# Engineering strain relaxation of GeSn epilayers on Ge/Si(001) substrates

Krista R Khiangte[1, *], Jaswant S Rathore[1, *], Vaibhav Sharma[1], Apurba Laha[2], Suddhasatta Mahapatra[1]

[1]Department of Physics, Indian Institute of Technology Bombay, MUMBAI 400076

[2]Department of Electrical Engineering, Indian Institute of Technology Bombay, MUMBAI 400076

*These authors contributed equally to this work

A mechanism of controlling the degree of strain relaxation in GeSn epilayers, grown by molecular beam epitaxy on Ge/Si(001) substrates, is reported in this work. It is demonstrated that by suitably controlling the thickness and the growth recipe of the underlying Ge buffer layer, both fully-strained and highly-relaxed GeSn epilayers can be obtained, without significant Sn segregation. The strain relaxation of the GeSn epilayer is mediated by threading dislocations of the Ge buffer layer, propagating across the Ge-GeSn interface. Systematic estimation of the threading dislocation density in both the alloy epilayer and the Ge buffer layer, by the approach developed by Benediktovitch et al. [A. Benediktovitch, A. Zhylik, T. Ulyanenkova, M. Myronov, A. Ulyanenkov (2015), J. Appl. Cryst. **48**, 655-665] supports this observation, and also reveals that no additional dislocations are generated at the Ge-GeSn interface. Together with recently reported techniques to arrest dislocation propagation in GeSn epilayers, these results bode extremely well for realization of highly-relaxed GeSn epilayers, much coveted for development of GeSn-based emitters.

Integration of photonic elements to the complementary-metal oxide-semiconductor (CMOS) platform is an important milestone for the semiconductor industry. In this context, the alloy of Ge and Sn (GeSn) has gained significant attention as a group-IV semiconductor with promising electronic [2,3] and optical properties [4,5]. Particularly encouraging for photonics is the fact that GeSn exhibits a direct energy bandgap, for Sn concentrations of ~ 6% or higher, provided epilayers of the alloy are fully relaxed [6-11]. However, light emitting devices with active layers of partially-relaxed direct-bandgap GeSn epilayers have already been fabricated [12, 13], including prototypes of optically-pumped lasers [14, 15].

For fully-strained or partially stain-relaxed GeSn epilayers, the indirect-to-direct bandgap transition is shifted towards higher Sn concentrations. Incorporation of such large concentrations of Sn in epitaxially grown GeSn/Ge/Si(001) heterostructures is challenging, due to the higher lattice-mismatch between the alloy and Ge, and the low solid-solubility of Sn in Ge [16]. Thus, one of the focus areas of research in GeSn epitaxy is to understand and engineer the relaxation behaviour of the alloy, such that high-quality epilayers with sufficiently high Sn-concentrations can be grown [17 – 22] for fabrication of efficient light-emitting devices. The usual approach to achieve complete strain relaxation is to increase the epilayer thickness much beyond the critical thickness for strain relaxation. In GeSn epitaxy, this however leads to poor surface quality and epitaxial breakdown, particularly for large Sn concentrations (> 10%). Moreover, achieving complete strain-relaxation, even in thick GeSn epilayers, appears to be a non-trivial task. For a layer thickness of 1 µm, von Driesch et al. observed only 81 % strain relaxation in chemical-vapour-deposition-(CVD)-grown GeSn epilayers, with 12.5 % Sn [17]. Subsequently, Margetis et al. achieved a much higher strain relaxation (95 %) in GeSn epilayers of similar thickness and Sn-composition [18], also grown by CVD. Interestingly, the authors observed a dislocation filtering effect, wherein threading dislocations formed half-loops within the first ~300 nm of the epilayer (consisting of ~ 8.7 % Sn) and thereby enabled the subsequent growth of a high-quality layer with enhanced Sn-content (12.7 %). Very recently, Aubin et al. demonstrated CVD-growth of step-graded GeSn epilayers, with 64 % strain relaxation of the topmost 180-nm layer, containing16 % Sn [19]. While these studies on strain relaxation of CVD-grown GeSn/Ge/Si(001) epilayers are very encouraging, it is important to note that such large degree of strain relaxation has not been widely reported in molecular beam epitaxy (MBE) growth of the alloy. Takeuchi et al. obtained up to 45 % strain relaxation in MBE-grown GeSn epilayers with 2.2 % Sn, as a result of post-growth thermal annealing for 10 min at 600 °C, in $N_2$ ambient [20]. However, by adopting the same approach for strain relaxation of GeSn epilayers with higher Sn-content (> 2.5 %), the authors observed β-Sn-precipitation and concomitant reduction of Sn-content from the bulk of the epilayer [21 - 22].

In this work, we show that strain-relaxation of GeSn epilayers grown on Ge/Si(001) by MBE can be tailored significantly, by controlling the crystal-quality and thickness of the underlying Ge buffer layer. Our results demonstrate that thin and partially-relaxed Ge buffer layers can induce strain-relaxation of the overgrown GeSn epilayer, while their thick and fully-relaxed counterpart, typically used in GeSn epitaxy, may supress the same. We achieved 72% and 78% strain relaxation in 450-nm-thick GeSn epilayers, with 5.4 % and 9.4% Sn, respectively.

The growth details of the three samples investigated in this work (labelled as A, B, and C) are given in Table 1. The Ge and the GeSn epilayers of all the three samples were grown on boron-doped silicon (001) substrates, in a RIBER C12 molecular beam epitaxy (MBE) chamber, maintained at a base pressure of $7 \times 10^{-10}$ mbar. The Ge buffer layer of sample A was grown by the two-step growth technique developed by Colace et al. [23], followed by two iterations of in-situ annealing [24]. In the two-step growth process, the first 20 nm of Ge was grown at $T_G$ = 250 °C, while the rest of the buffer layer was grown at $T_G$ = 400 °C. In-situ annealing was performed at $T_A$ = 840 °C for $t_A$ = 10 mins, interspersed by a cooling-down step (to 250 °C). The two-step growth technique, together with the post-growth cyclic annealing treatment, is a well-established route to obtain Ge buffer layers of high crystalline quality, for layer thicknesses of ~ 300 nm or higher. The recipe followed in sample A for the Ge buffer growth has been optimized [25] for the most superior crystal quality, as demonstrated later by HRXRD and HRTEM results. On the contrary, the Ge buffer layers for samples B and C were intentionally grown at a low temperature ($T_G$ = 250 °C) and with a smaller thickness (80 nm). In low temperature Ge epitaxy, the misfit strain is predominantly released plastically, by formation of dislocations. These dislocations are known to thread through the Ge layer, for the small thicknesses chosen for samples B and C. For all three samples, the GeSn epilayer was grown at $T_G$ = 180 °C, at a growth rate of 1.4 nm min$^{-1}$. Further details regarding pre-growth surface preparation and growth of the epilayers can be found in Ref. [25].

HRXRD has been extensively employed in this work to determine the strain-state of both the GeSn and the Ge epilayers, and the Sn-content of the former. More importantly, the threading dislocation densities (TDD) of both the layers have also been estimated by HRXRD, following the method proposed very recently by Benediktovich et el. [1]. This approach is a generalization of the technique originally developed by Kaganer et al. (for c-oriented GaN/Sapphire (0001) epilayers [26]) for arbitrary surface-orientation and dislocation-line-direction and provides a better estimate of the TDD, in comparison to the usual methods [25, 27-29] which rely on a Gaussian (or Voigt) fitting of the diffracted X-ray intensity. All X-rays measurements were carried out in a Rigaku Smartlab diffractometer, equipped with a 9 kW rotating Cu anode, a parabolic mirror, and a double-crystal Ge (220) monochromator, and capable of performing scans in both out-of-plane and in-plane geometries. The HRXRD set-up for an arbitrary asymmetric scan is shown schematically in Figure 1 (a), wherein the important angles of the diffraction geometry are depicted. In this work, $\omega - 2\theta$ scans and $\omega$- scans have been recorded for the symmetric (004) and the asymmetric (224) reflections. Diffractograms were recorded in the double-crystal configuration, with a wide open detector. The crystalline quality of the epilayers was further probed by cross-sectional high-resolution transmission electron microscopy (XTEM), using a JEOL 200 microscope operating at voltages of up to 200 kV.

*Table 1: Layer-thicknesses ($t_{Ge}$ and $t_{GeSn}$) and growth temperatures ($T_{Ge}$ and $T_{GeSn}$) of different samples studied in this work. The Ge buffer layer of sample A was further (cyclic-) annealed (See main text).*

| Sample | Ge buffer layer | | GeSn epilayer | |
|---|---|---|---|---|
| | $T_{Ge}$ (°C) | $t_{Ge}$ (nm) | $T_{GeSn}$ (°C) | $t_{GeSn}$ (nm) |
| A | 250 (~ 20 nm) + 400 (~330 nm) | 352 | 180 | 500 |
| B | 250 | 80 | 180 | 450 |
| C | 250 | 80 | 180 | 450 |

Figure 1 (b) and 1 (c) show the $\omega - 2\theta$ diffractograms of the (004) and (224) reflections, for all three samples. In case of sample A, it is observed that the Ge (004) reflection exhibits an asymmetric broadening towards the Si (004) reflection (Fig. 1 (b) (top panel)). This suggests the presence of an alloyed SiGe layer at the Si-Ge interface. Si-Ge intermixing is a usual occurrence with post-growth high-temperature cyclic annealing [23]. The in-plane and out-of-plane lattice constants of both Ge and GeSn epilayers, as calculated from the positions of the intensity maxima of the (004) and (224) reflections, are listed in Table 2.

Also listed are the bulk lattice constants and the Sn-concentrations of the GeSn epilayers, as estimated using Vegard's law [30]. It is worth noting that the GeSn (004) reflection of sample C is also asymmetrically broadened, towards lower $2\theta$ values. This suggests that the Sn-content varies across the thickness of the GeSn epilayers, which is most likely due to the tendency of Sn to segregate to the surface. In calculating the bulk lattice constant and the Sn-content of the GeSn alloy in sample C, the position of the intensity-maximum has been considered. It may therefore be concluded that the tabulated value corresponds to the minimum Sn-content (9.4%) of the alloy epilayer in sample C.

*Table 2*: *The measured values of in-plane and out-of-plane lattice constants of the Ge and GeSn layers, together with the calculated values of bulk lattice constants and Sn contents of the alloy.*

| Sample | Sn (%) | $a_\perp^{Ge_{1-x}Sn_x}$ (Å) | $a_\parallel^{Ge_{1-x}Sn_x}$ (Å) | $a^{Ge_{1-x}Sn_x}$ (Å) | $a_\perp^{Ge}$ (Å) | $a_\parallel^{Ge}$ (Å) | $a^{Ge}$ (Å) |
|---|---|---|---|---|---|---|---|
| A | 6.2 | 5.744 | 5.667 | 5.726 | 5.648 | 5.666 | 5.652 |
| B | 5.4 | 5.720 | 5.683 | 5.701 | 5.662 | 5.636 | 5.647 |
| C | 9.4 | 5.756 | 5.713 | 5.734 | 5.676 | 5.638 | 5.656 |

The thicknesses of the GeSn epilayers ($t_{GeSn}$) are plotted in Figure 1 (d) versus their Sn-concentrations. Also plotted in the same Figure are the expected critical thicknesses ($t_c$) for onset of plastic relaxation, as predicted by the Mathew-Blakeslee [31] and the People-Bean [32, 33] models. For all three samples, $t_{GeSn}$ is larger than the $t_c$ predicted by either of the Mathew-Blakeslee (M-B) and People-Bean (P-B) models, with the deviation being much larger in case of the former. However, the systematic study of Ref. [34] conclusively demonstrates that the P-B model provides the more realistic estimate of $t_c$, in case of GeSn epitaxy. With respect to the P-B model, $t_{GeSn}/t_c$ is nearly the same for samples A and B (1.72 and 1.12, respectively), while it is very different for sample C ($t_{GeSn}/t_c = 4.1$). Therefore, the degree of strain relaxation ($R$) may be expected to be comparable for samples A and B, while being significantly higher for sample C. Furthermore, one would also expect $R$ for sample A to be higher than that for sample B.

Interestingly, the plot of $R_{Ge_{1-x}Sn_x}$ for the three samples, depicted in Figure 1 (e), do not corroborate to this intuitive correlation. Here, $R_{Ge_{1-x}Sn_x}$ has been calculated as

$$R_{Ge_{1-x}Sn_x} = \left(a_{||}^{Ge_{1-x}Sn_x} - a_{||}^{Ge}\right)/\left(a^{Ge_{1-x}Sn_x} - a_{||}^{Ge}\right) \quad (1)$$

Also plotted in Fig. 1 (e) is the degree of strain relaxation of the Ge buffer layers, calculated as $R_{Ge} = \left(a_{||}^{Ge} - a^{Si}\right)/\left(a^{Ge} - a^{Si}\right)$. It is seen that while in sample A, the GeSn epilayer is nearly pseudomorphically strained ($R_{Ge_{0.938}Sn_{0.062}} = 1.5\,\%$), the alloy epilayer of sample B is significantly relaxed ($R_{Ge_{0.966}Sn_{0.054}} \approx 72\,\%$). In fact, $R_{Ge_{0.966}Sn_{0.054}}$ corresponding to sample B is rather comparable to that of sample C ($R_{Ge_{0.906}Sn_{0.094}} \approx 78\%$). Thus, it is evident that the strain relaxation of the GeSn epilayers in this study is not driven by the plastic relaxation of misfit-induced strain at the Ge-GeSn interface.

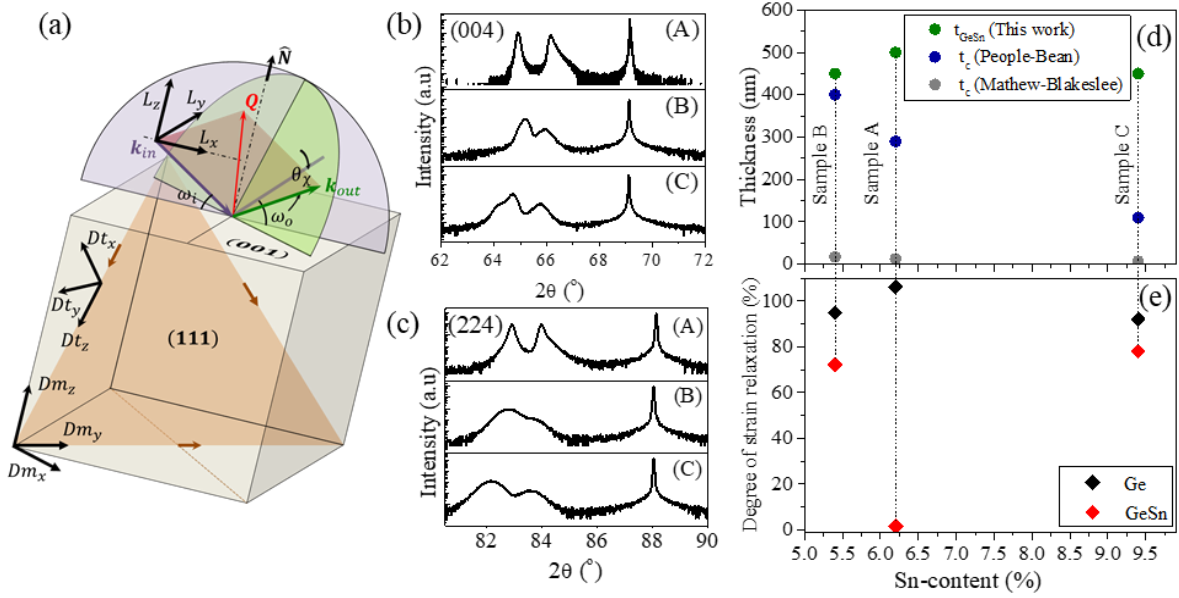

*Figure 1: (a) A schematic representation of the diffraction geometry and (section of) the slip system of the fcc crystal structure. $k_i$ and $k_o$ are the incident and the diffracted wave vectors, respectively, while $\widehat{N}$ and $Q$ represent the normal to the sample surface and the diffraction vector, respectively. $\omega_i$ is the angle between the sample surface and the line connecting the source to the sample plane. $\omega_o$ is the angle between the sample plane and the projection of $k_o$ onto the plane defined by $k_i$ and $\widehat{N}$. The slip system of the fcc crystal structure is represented by $\left(\frac{a}{2}\langle 110\rangle\{111\}\right)$. On the (111) plane, the possible slip directions are shown by the arrows. Three different co-ordinate systems, $(L_x, L_y, L_z)$, $(Dm_x, Dm_y, Dm_z)$, and $(Dt_x, Dt_y, Dt_z)$, used in the dislocation density analysis, are shown in the schematic. (b) and (c) $\omega - 2\Theta$ diffractograms for the symmetric (004) and the asymmetric (224) reflections, collected from the series of samples. In all cases, the GeSn peaks are to the left of the Ge peaks, while the Si*

*substrate peaks are the right-most. (d) A plot of the sample thicknesses, together with the critical thicknesses for strain relaxation for the GeSn-Ge heterosystems, as predicted by the Mathew-Blakeslee and the People-Bean models. (e) A plot of the degree of strain relaxation of both the Ge and the GeSn layers for all three samples versus the Sn-content of the GeSn epilayers.*

To understand this counter-intuitive observation, it is important to consider the strain-state and the thicknesses of the Ge buffer layers as well. In sample A, the Ge buffer layer is fully-relaxed (in fact, it is tensile-strained due to thermal-conductivity-mismatch induced residual strain [35, 36]) and 350 nm thick, whereas in sample B and C, it is 94.9 % and 92 % relaxed, respectively, and only 80 nm thick. As will be demonstrated in the sections below, these attributes of the Ge buffer layer play the most crucial role in controlling the strain relaxation of the alloy epilayer atop. A large density of dislocations thread through the Ge-GeSn interface in case of the partially-relaxed thin Ge buffer layers, which cause the GeSn epilayers to relax (irrespective of the alloy composition). In the thick Ge buffer layers, grown by the two-step growth technique (and cyclic annealing), dislocation threading is arrested, which in turn, supresses the relaxation of the GeSn epilayer.

To further determine the role of the Ge buffer layer in the relaxation of the GeSn epilayer, quantitative estimates of the TDD were obtained from the ω-diffractograms of the (004) reflections, for both the GeSn and the Ge layers, by the Benediktovich approach. As outlined in Ref. [1], the measured intensity distribution of the rocking curve, in the direction of the scattered beam ($\hat{\boldsymbol{n}} = \boldsymbol{k}_{out}/k_{out}$) is given by

$$I(q_n)_M = I(\boldsymbol{q}.\hat{\boldsymbol{n}}) = \int_{-\infty}^{\infty} d\delta x \int_0^d dz \exp(i\boldsymbol{q}.\hat{\boldsymbol{n}}\delta x)\, G(\delta x, z) \qquad (2)$$

The term $G(\delta x, z)$ is the correlation function, expressed as

$$G(\delta x, z) = \exp\bigl(-iT_1(\delta x) - T_2(\delta x, z)\bigr) \qquad (3)$$

where $T_1(\delta x) = \rho_m Q_i \langle \epsilon_{ij}^{(L)} \rangle n_j \delta x$ and $T_2(\delta x, z) = \frac{1}{2}\rho Q^2 G_{ijkl}^{(L)} E_{ijkl}^{(L)} \delta x^2$. Here, $\boldsymbol{Q}$ is the scattering vector (See Fig. 1 (a)) and $\boldsymbol{q} = (\boldsymbol{Q} - \boldsymbol{Q}_0)$ describes the deviation of the scattering vector from the reciprocal lattice vector $\boldsymbol{Q}_0$. The tensors $\langle \epsilon_{ij}^{(L)} \rangle$ and $E_{ijkl}^{(L)}$ describe the mean strain due to misfit dislocations and the strain fluctuation, respectively, both represented in the "laboratory" (L) co-ordinate system, $L_z \| \widehat{\boldsymbol{N}}$ and $L_x \| [\boldsymbol{k}_{in} - \widehat{\boldsymbol{N}}(\widehat{\boldsymbol{N}}.\boldsymbol{k}_{in})]$ (See Fig. 1 (a)). $\widehat{\boldsymbol{N}}$ is

the normal to the sample surface and $\boldsymbol{k}_{in}$ denotes the incoming wave vector. Finally, the term $G_{ijkl}^{(L)}$ is the geometric tensor, which captures the influence of the measurement mode on the measured intensity distribution. We note here that the expression for $G(\delta x, z)$ used in Ref. [1] is incorrect and has been corrected here in Equation (3).

The term $\rho_m$ denotes the density of misfit dislocations, while the expression of $\rho$ contains both $\rho_m$ and the threading dislocation density ($\rho_s$) as,

$$\rho E_{ijkl}^{(L)} = \frac{g\rho_m}{d} \sum_\alpha T_{\alpha ijkl}^{LD\,i'j'k'l'} E_{i'j'k'l'}(z)^{(Dm)} + \rho_s \sum_\alpha T_{\alpha ijkl}^{LD\,i'j'k'l'} E_{i'j'k'l'}(z)^{(Dt)} \qquad (4)$$

Here, $(Dm)$ and $(Dt)$ denote the co-ordinate systems described by $\left(Dm_z \| \widehat{\boldsymbol{N}}, Dm_y \| \widehat{\boldsymbol{\tau}}\right)$ and $\left(Dt_z \| \widehat{\boldsymbol{\tau}}, Dt_{x,y} \perp \widehat{\boldsymbol{\tau}}\right)$ (See Fig. 1 (a)), while $T_{\alpha ijkl}^{LD\,i'j'k'l'}$ are the co-ordinate transformation tensors. The summation index $\alpha$ denotes the different dislocation line directions and $g$ is a positional correlation parameter for the misfit dislocations. The final expression for the diffraction intensity simplifies to

$$I(\boldsymbol{q}.\widehat{\boldsymbol{n}}) = \int_0^d \sqrt{\frac{\pi}{T_2(z)}} \exp\left(-\frac{\boldsymbol{q}.\boldsymbol{n} - \rho_m Q_i \langle \epsilon_{ij} \rangle n_j}{4T_2(z)}\right) dz \qquad (5)$$

For each dislocation line direction $\alpha$, corresponding to the slip system appropriate for Ge and GeSn (fcc, $\left(\frac{a}{2}\langle 110\rangle\{111\}\right)$), the strain tensors were calculated (following the corresponding expressions in Ref. [1]). The fit of expression (5) to the measured ω-diffractograms of both the GeSn and the Ge layers shows an excellent agreement, even down to the low intensity tails, as shown in Fig. 2 (a) and 2(b). The obtained values of the threading dislocation densities (TDD), for the Ge and the GeSn epilayers, are plotted in Figure 2 (c). There are two important features to note in this plot. Firstly, for all three samples, the TDDs are comparable in the Ge buffer and the GeSn epilayer, with the latter being marginally smaller. Secondly, the TDD measured for sample A is about two-orders-of-magnitude lower than that of samples B and C. Both observations indicate that the dislocations causing the GeSn layers to relax are those which thread from the Ge buffer layer underneath. The marginal reduction in the TDD of the GeSn epilayers, in comparison to the Ge buffer layers is probably related to the interaction of threading dislocations, mediated by Sn atoms in the alloy epilayer [37].

Cross-sectional transmission electron microscopy (XTEM) images of the three samples are shown in Figure 3. For sample A, the dislocation network of the Ge buffer layer is seen to be confined in a region close to the Si-Ge interface (Fig. 3 (a)). The dark-field image (Fig. 3 (b)) shows annihilation of 60° dislocations within ~ 100 nm of the buffer layer thickness.

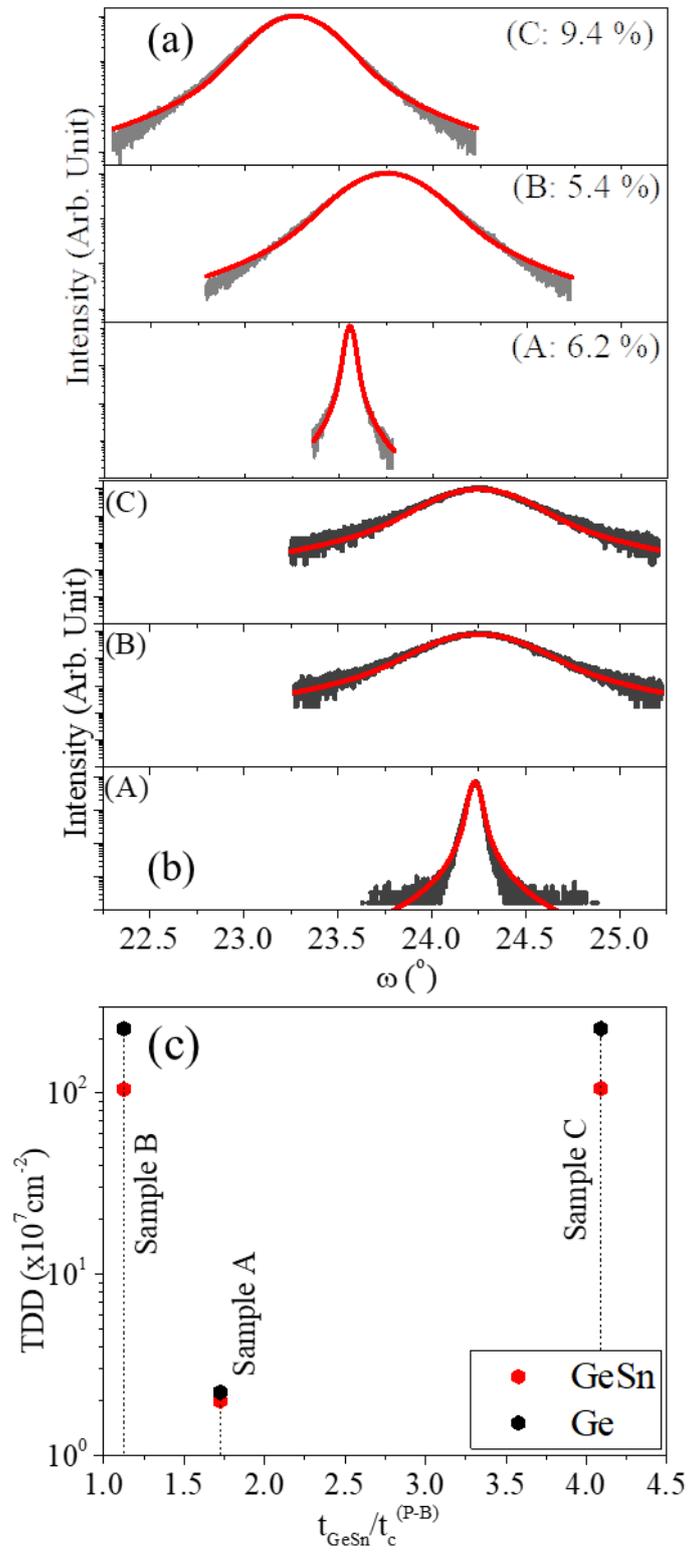

*Figure 2: Fitting of expression (5) to (004) ω- diffractograms corresponding to the (a) GeSn epilayers and (b) Ge buffer layers of the three samples (c) Plot of the threading dislocation densities (TDD) for both the Ge and the GeSn layers versus the ratio $t_{GeSn}/t_c$, where $t_c$ is as predicted by the People-Bean model.*

Confinement of dislocations at the epilayer/substrate interface is a characteristic outcome of cyclic annealing in Ge/Si epitaxial layers and consistent with earlier reports [24, 38]. Concomitantly, dislocation threading across the Ge-GeSn interface is barely visible. On the other hand, for samples B and C, a large density of threading dislocations is observed to cross the Ge-GeSn interface, from the thin, partially relaxed Ge buffer layer (Figs. 3 (c) and 3 (d), respectively). As argued earlier in the context of the HRXRD results, these dislocations cause the GeSn epilayer to relax, even observed for sub-critical layer thicknesses [25].

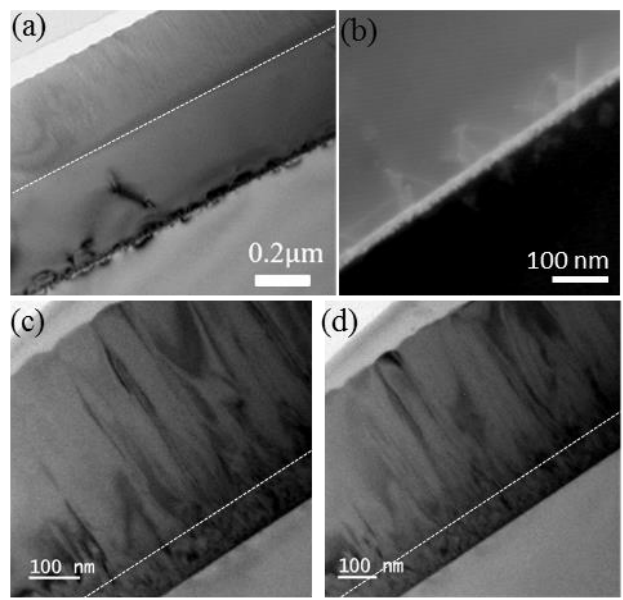

*Figure 3: XTEM bright field images recorded for samples (a) A (6.2% Sn), (c) B (5.4% Sn) and (d) C (9.4% Sn), and the dark field image recorded for sample A (b). The white dotted lines show the interface between the Ge and the GeSn layers.*

It remains to investigate how the propagation of the threading dislocations may be suppressed within the GeSn epilayer, such that beyond a reasonable layer thickness, high quality epitaxy may be recovered. Sn atoms have been demonstrated in previous reports to induce interaction of threading dislocations, to yield pure edge-type (Lomer) dislocations [39]. This fact may be advantageously exploited by introducing pure Sn in sub-monolayer quantity, after growth of a few tens of nanometer of the alloy epilayer. The efficacy of this approach needs further experimental investigations.

In conclusion, we have proposed a possible mechanism of strain relaxation in GeSn epilayers, by tailoring the growth recipe and thickness of the underlying Ge buffer layer, even for sub-critical layer thicknesses. Combined with additional techniques of dislocation filtering, this approach may pave the way for obtaining highly-relaxed GeSn epilayers by MBE, without significantly compromising the crystalline quality.

**Acknowledgements**

The research was funded by the Science and Engineering Research Board (SERB), Department of Science and Technology (DST), Government of India. The authors acknowledge support from the Centre of Excellence in Nanoelectronics (CEN) and Industrial Research and Consultancy Centre (IRCC), of the Indian Institute of Technology Bombay (IITB).

**References:**

1. A. Benediktovitch, A. Zhylik, T. Ulyanenkova, M. Myronov, A. Ulyanenkov, J. Appl. Cryst. 48 (2015) 655.
2. Y. C. Fang, K. Y. Chen, C. H. Hsieh, C. C. Su, Y. H. Wu, ACS Appl. Mater. Interfaces 7(48) (2015) 26374.
3. G. Han, Y. Wang, Y. Liu, C. Zhang, Q. Feng, M. Liu, S. Zhao, B. Cheng, J. Zhang, Y. Hao, IEEE Electron Device Lett. 37 (2016) 701.
4. R. Soref, D. Buca, S.-Q. Yu, Opt. and Photon. News 27(2016) 32.
5. J. Mathews, R. Roucka, J. Xie, S. Yu, J. Menéndez, J. Kouvetakis, Appl. Phys. Lett. 95 (2009) 133506
6. J. Kouvetakis, J. Menendez, A.V.G. Chizmeshya, Annu. Rev. Mater. Res. 36 (2006) 497
7. G. He, H.A. Atwater, Phys. Rev. Lett. 79 (1997) 1937.
8. W. -J. Yin, X.-G. Gong, S.-H. Wei, Phys. Rev. B 78 (2008) 161203.
9. V. R. D'Costa, C. S. Cook, A. G. Birdwell, C. L. Littler, M. Canonico, S. Zollner, J. Kouvetakis, J. Menendez , Phys. Rev. B 73 (2006) 125207.
10. H. Lin, R. Chen, W. Lu, Y. Huo, T. I. Kamins, J. S.Harris, Appl. Phys. Lett. 100 (2012) 102109.
11. J. Mathews, R. T. Beeler, J. Tolle, C. Xu, R. Roucka, J. Kouvetakis, J. Menendez, Appl. Phys. Lett. 97 (2010) 221912.
12. H. H. Tseng, K. Y. Wu, H. Li, V. Mashanov, H. H. Cheng, G. Sun, R. A. Soref, Appl. Phys. Lett. 102 (2013) 182106.
13. Y. Zhou, W. Dou, W. Du, T. Pham, S. A. Ghetmiri, S. Al-Kabi, A. Mosleh , M . Alher, J. Margetis, J. Tolle, G. Sun, R. Soref , B. Li , M. Mortazavi, H. Naseem, S.-Q. Yu, J. Appl. Phys. 120 (2016) 023102.
14. S. Wirths, R. Geiger, N. von den Driesch, G. Mussler, T. Stoica, S. Mantl, Z. Ikonic, M. Luysberg, S. Chiussi, J. M. Hartmann, H. Sigg, J. Faist, D. Buca, D. Grützmacher, Nature Photon. 4 (2015) 88.
15. D. Stange, S. Wirths, R. Geiger, C. Schulte-Braucks, B. Marzban, N. von den Driesch, G. Mussler, T. Zabel, T. Stoica, J. M. Hartmann, S. Mantl, Z. Ikonic, D. Grützmacher, H. Sigg, J. Witzens, D. Buca ACS Photon. 3 (2016) 1279.
16. E. Kasper, J. Werner, M. Oehme, S. Escoubas, N. Burle, J. Schulze Thin Solid Films 520 (2012) 3195.

17. N. von den Driesch, D. Stange, S. Wirths, G. Mussler, B. Holländer, Z. Ikonic, J. M. Hartmann, T. Stoica, S. Mantl, D. Grützmacher, D. Buca, Chem. Mater. 27 (2015) 4693.
18. J. Margetis, A. Mosleh, S. Al-Kabi, S. A. Ghetmiri, W. Du, W. Dou, M. Benamara, B. Li, M. Mortazavi, H.A. Naseem, S.-Q. Yu, J. Tolle, J. Crystt. Growth 463 (2017) 128.
19. J. Aubin, J. M. Hartmann, A. Gassenq, J. L. Rouviere, E. Robin, V. Delaye, D. Cooper, N. Mollard, V. Reboud, V. Calvo, Semicond. Sci. Technol. 32 (2017) 094006.
20. S.Takeuchi, A. Sakai, K. Yamamoto, O. Nakatsuka, M. Ogawa, S. Zaima, Semicond. Sci. Technol. 22 (2007) S231
21. S. Takeuchi, Y. Shimura, O. Nakatsuka, S. Zaima, M. Ogawa, A. Sakai Appl. Phys. Lett. 92, (2008) 231916.
22. T. Asano, S. Kidowaki, M. Kurosawa, N. Taoka, O. Nakatsuka, S. Zaima, Thin Solid Films 557 (2014) 159.
23. L. Colace, G. Masini, F. Galluzzi, G. Assanto, G. Capellini, L. D. Gaspare, E. Palange, F. Evangelisti, Appl. Phys. Lett. 72 (1998) 3175.
24. H. Luan, D. R. Lim, K. K. Lee, K. M. Chen, J. G. Sandland, K. Wada, L. C. Kimerling, Appl. Phys. Lett. **75** (1999) 2909.
25. K. R. Khiangte, J. S. Rathore, V. Sharma, S. Bhunia, S. Das, R. S. Fandan, R. S. Pokharia, A. Laha, S. Mahapatra, J. Cryst. Growth 470 (2017) 135.
26. V. M. Kaganer, R. Köhler, M. Schmidbauer, R. Opitz, B. Jenichen Phys. Rev. B. 55 (1997) 1793.
27. J. E. Ayers J. Cryst. Growth 135 (1994) 71.
28. G.K. Williamson, W.H. Hall Acta Metall. 1 (1953) 22.
29. C.G. Dunn, E. F. Koch, Acta Metall. 5 (1957) 548.
30. R. Cheng, W. Wang, X. Gong, L. Sun, P. Guo, H. Hu, Z. Shen, G. Han and Y. -C Yeo, ECS J. Solid State Sci. Technol. 2 (2013) 138.
31. J. W. Matthews, A.E. Blakeslee, J. Cryst. Growth 27 (1974) 118.
32. R. People, J. C. Bean, Appl. Phys. Lett. 47 (1985) 322.
33. R. People, J. C. Bean, Appl. Phys. Lett. 49 (1986) 229.
34. W. Wang, Q. Zhou, Y. Dong, E. S. Tok, Y. -E. Yeo, Appl. Phys. Lett. 106 (2015) 232106.
35. Y. Ishikawa, K. Wada, D. D. Cannon, J. Liu, H. Luan, L. C. Kimerling, Appl. Phys. Lett. 82 (2003) 2044.

36. D. D. Cannon, J. F. Liu, Y. Ishikawa, K. Wada, D. T. Danielson, S. Jongthammanurak, J. Michel, L. C. Kimerling, Appl. Phys. Lett. 84 (2004) 906.
37. A. Mosleh, M. Benamara, S. A Ghetmiri, B. R. Conley, M. A. Alher, W. Du, G. Sun, R. Soref, J. Margetis, J. Tolle, S.-Q. Yu, H. A. Naseem, ECS Trans. 64 (2014) 1845.
38. Z. Liu, X. Hao, A. H. Baillie, Ch. –Y. Tsao, M. A. Green, Thin Solid Films, 574 (2015) 99.
39. A. F. Marshall, D. B. Aubertine, W. D. Nix, P .C. McIntyre, J. Mater. Res. 20 (2005) 447.